\RequirePackage{ifpdf}
\ifpdf 
\documentclass[pdftex]{sigma}
\else
\documentclass{sigma}
\fi
\usepackage{cite}
\usepackage{dsfont}
\usepackage{tikz-cd}
\usepackage{enumitem}

\usepackage{mathrsfs} 

\numberwithin{equation}{section}

\numberwithin{theorem}{section}
\numberwithin{proposition}{section}
\numberwithin{lemma}{section}
\numberwithin{corollary}{section}
\numberwithin{definition}{section}
\numberwithin{example}{section}
\numberwithin{remark}{section}
\numberwithin{note}{section}

\begin{document}
\begin{flushright}
\Large{\textmd{DESY--21--158}}
\end{flushright}

\renewcommand{\PaperNumber}{***}

\FirstPageHeading

\ShortArticleName{Mellin-Barnes integrals related to the Lie algebra~$u(N)$}

\ArticleName{Mellin - Barnes integrals related to the Lie algebra~$u(N)$}

\Author{ Alexander N.~MANASHOV}

\AuthorNameForHeading{ A.N.~Manashov}

\Address{Institut f\"ur Theoretische Physik, Universit\"at Hamburg, D-22761 Hamburg, Germany,
\\
Saint - Petersburg Department of Steklov Mathematical Institute of Russian Academy of Sciences, Fontanka 27, 191023 Saint - Petersburg,
Russia. } \EmailD{\href{mailto:alexander.manashov@desy.de}{alexander.manashov@desy.de}}


\Abstract{We present an alternative proof of Gustafson's generalization\\
 of the second  Barnes' lemma.}



\section{Introduction}\label{sect:introduction}

In refs.~\cite{Gustafson92,Gustafson94} R.~A.~Gustafson  generalized the first and the second Barnes' lemmas to
the case of the  Lie algebra
$u(n)$. Namely, he calculated the following  multidimensional Mellin~-~Barnes (MB) integrals  in closed form:
\begin{align}\label{Gustafson-I}
    \int_{-i\infty}^{i\infty}\cdots \int_{-i\infty}^{i\infty}
        \frac{\prod_{k=1}^{N+1} \prod_{j=1}^{N} \Gamma(\alpha_k-z_j)\Gamma(\beta_k+z_j)}{
        \prod_{1\leq k< j\leq N}\Gamma(z_k-z_j\Gamma(z_j-z_k)} \prod_{k=1}^N \frac{dz_k}{2\pi i} =
=\frac{N!\prod_{k,j=1}^{N+1} \Gamma(\alpha_k+\beta_j)}{\Gamma\big(\sum_{k=1}^{N+1}(\alpha_k+\beta_k )\big)}
\end{align}
and
\begin{flalign}\label{Gustafson-IA}
    \int_{-i\infty}^{i\infty}\cdots \int_{-i\infty}^{i\infty}
        \frac{ \prod_{j=1}^{N} \prod_{k=1}^{N+2}\Gamma(\alpha_k-z_j)\prod_{m=1}^{N+1}\Gamma(\beta_m+z_j)}{
        \prod_{m=1}^{N}\Gamma(\gamma-z_m)
        \prod_{1\leq k< j\leq N}\Gamma(z_k-z_j\Gamma(z_j-z_k)} \prod_{k=1}^N \frac{dz_k}{2\pi i}
        &=
\frac{N!\prod_{k=1}^{N+2} \prod_{j=1}^{N+1} \Gamma(\alpha_k+\beta_j)}{\prod_{k=1}^{N+2}\Gamma(\gamma-\alpha_k )}. &
\end{flalign}
Here $\gamma = \sum_{k=1}^{N+2} \alpha_k+\sum_{k=1}^{N+1} \beta_k$ and it is assumed that the integration contours separate  sequences of
poles going to the right, $\alpha_n + k$, and to the left, $-\beta_m - k$, $n=1,\ldots,N+1(N+2)$, $m=1,\ldots,N+1$, which are  due to the
gamma--functions in the numerators.

The integral~\eqref{Gustafson-IA}  depends on $2n+3$ external parameters and  implies the relation in~\eqref{Gustafson-I}.
Indeed, sending $\alpha_{N+2}\to \infty$ in \eqref{Gustafson-IA} one recovers~\eqref{Gustafson-I}. The  integral~\eqref{Gustafson-IA}
was calculated in ref.~\cite{Gustafson94} making use of the residues theorem and evaluating the corresponding sums with
the help of Milne's $\mathrm{U}(n)$
generalization of the Gauss summation theorem~\cite{Milne88}.
The proof of the MB integral~\eqref{Gustafson-IA} given in \cite{Gustafson92}
is more involved and follows a different route. It relies on the integral~\eqref{Gustafson-I} and uses induction on $N$ to show that the
expressions on the l.h.s. and the r.h.s. of \eqref{Gustafson-IA} coincide for special values of the external parameters.
The general case then follows from Carlson's theorem.
In the present paper we present an alternative non-inductive derivation of the integral~\eqref{Gustafson-IA}.

Classical MB integrals can be extended to the $q$-beta and elliptic integrals, see
refs.~\cite{Gustafson92,Gustafson94,SpiridonovEssays,SpiridonovWarnaar,Rains}, and to the MB integrals which involve the  gamma--functions
over the field of complex numbers~\cite{MR2125927}. In the latter case it was shown~\cite{DM2020} that studying properties of the
counterpart of the integral~\eqref{Gustafson-I} as a function of external parameters one can  recover the corresponding analogue of the
second integral~\eqref{Gustafson-IA}. We want to apply the same approach, with some modifications, to the classical MB integrals  with
Euler's gamma functions. However, the integral~\eqref{Gustafson-I} is not a good starting point for such an analysis so we first consider a
different MB integral with more suitable analytic  properties.

\section {Auxiliary integral}

Let us consider the following MB integrals
\begin{flalign}\label{mod-Gust}
\mathscr R_\pm(a,\alpha,\beta) &= \int_{-i\infty}^{i\infty}\cdots \int_{-i\infty}^{i\infty} e^{\pm i\pi \sum z_k}
 \frac{\prod_{j=1}^{N}\left(\prod_{k=1}^{N+1}\Gamma(\alpha_k-z_j) \prod_{m=1}^N\Gamma(z_j+\beta_m)\right)}{\prod_{k=1}^N
\Gamma({a}-z_k) \prod_{k < j} \Gamma(z_k-z_j) \Gamma(z_j-z_k)}\prod_{k=1}^N \frac{dz_k}{2\pi i}. &
\end{flalign}
These integrals are close cousins of the integral \eqref{Gustafson-I}: the only difference is that we moved one gamma function from the
numerator to the denominator, $\Gamma(z_k+\beta_{N+1})\to1/\Gamma(a-z_k)$ and added the exponential  factor, ${e^{\pm i\pi \sum z_k}}$.
This factor compensates some sign factors arising during an evaluation of the integral by the residues theorem and changes its analytic
properties as a function of the external parameters, $\{a,\alpha_k,\beta_j\}$. Indeed, while the integral~\eqref{Gustafson-I} converges (at
large $z_k$) for all values of the external parameters, the convergence of the integrals~\eqref{mod-Gust} is controlled by the parameter
\begin{align}
\nu=a-\sum_{k=1}^{N+1}\alpha_k-\sum_{k=1}^{N}\beta_k.
\end{align}
 The integrals $\mathscr R_\pm$ converge only if $\text{Re}(\nu)>0$.

Assuming that this condition is fulfilled,  the integrals~\eqref{mod-Gust}  can be calculated according to the strategy used in
ref.~\cite{Gustafson94} for  the integral~\eqref{Gustafson-I}.  Namely, one closes the integration contours in the
left-half plane picking up the residues at  the poles of the gamma functions which are located
at $\{-\beta_{j} -k,\  k=0,1,\ldots,\  1\leq j\leq N\}$. Taking into
account the symmetry of the integrand under permutations of the arguments one derives in this way
\begin{flalign}\label{residues}
\mathscr R_\pm(a,\alpha,\beta)&=N!e^{\mp i\pi \mathscr B}\sum_{n_1,\ldots,n_N=0}^\infty \frac{1}{n_1!\ldots n_N!}
\prod_{k < j}\frac{\sin\pi(\beta_j-\beta_k)}\pi(-1)^{n_j+n_k}(\beta_k+n_k-\beta_j-n_j)\times
\notag\\
&\quad
\prod_{j=1}^{N}\frac{\prod_{k=1}^{N+1}\Gamma(\alpha_k+\beta_j+n_j)}{\Gamma(a+\beta_j+n_j)}
\prod_{1\leq k\neq j\leq N}^N \Gamma(\beta_k-\beta_j-n_j) &
\notag\\
&=N!e^{\mp i\pi \mathscr B}\frac{\prod_{j=1}^{N}\prod_{k=1}^{N+1}\Gamma(\alpha_k+\beta_j)}{\prod_{j=1}^{N}\Gamma(a+\beta_j)}\times
\notag\\
&\quad
\sum_{n_1,\ldots,n_N=0}^\infty \prod_{k < j}\frac{\beta_k+n_k-\beta_j-n_j}{\beta_k-\beta_j}
\prod_{j=1}^{N}\prod_{k=1}^{N+1}\frac{(\alpha_k+\beta_j)_{n_j}}{(1-\beta_k+\beta_j)_{n_j}}.
\end{flalign}
Here $\beta_{N+1}\equiv 1-a$ and $(a)_n$ stands for the Pochhammer symbol.  Finally, evaluating the sum in the last line of
Eq.~\eqref{residues} with the help of Milne's generalization of the Gauss summation theorem, see e.g.~\cite[Theorem~2.7]{MR911651}, one
obtains
\begin{align}\label{Rpm1}
\mathscr R_\pm(a,\alpha,\beta)& = N!\, e^{\mp i\pi \mathscr B}\,\Gamma(a-\mathscr B-\mathscr A)
\frac{\prod_{k=1}^{N}\prod_{j=1}^{N+1}\Gamma(\alpha_j+\beta_k)}{\prod_{j=1}^{N+1}\Gamma(a-\alpha_j)},
\end{align}
where $\mathscr B=\sum_{k=1}^N\beta_k$ and $\mathscr A=\sum_{j=1}^{N+1}\alpha_j$.  Note that the pole of the first gamma function at $\nu=
a-\mathscr B-\mathscr A=0$ signals that the  integrals diverge for these values of the parameters.
It is quite remarkable that the residues
of $\mathscr R_\pm$ at $\nu=0$ can be itself represented as certain MB integrals. In order to find these integrals it is convenient to
represent the original integrals~\eqref{mod-Gust} in  a determinant form.

\section{Determinant representation for $\mathscr R_\pm$}
Taking into account that
\begin{align}
\prod_{k<j}\frac1{\Gamma (z_{k}-z_{j})\Gamma( z_{j}-z_{k})}=\prod_{k<j}\frac1\pi {(z_j-z_k)}\,{\sin\pi (z_k-z_j)}
\end{align}
one gets the  following representation for $\mathscr R_\pm$
\begin{multline}\label{RR2}
\mathscr R_\pm(a,\alpha,\beta)=\pi^{-N(N-1)/2}\frac1{(2\pi i)^N}\int_{-i\infty}^{i\infty}\cdots\int_{-i\infty}^{i\infty}
\prod_{k=1}^N  \big(\cos\pi z_k\big)^{N-1}\times
\\
\prod_{k<j} (z_k-z_j) \Big(\tan \pi z_j-\tan\pi z_k\Big)\,
Q_\pm(z_1)\ldots Q_\pm(z_N) dz_1\dots dz_N,
\end{multline}
where
\begin{align}
Q_\pm(z) = \frac{e^{\pm i\pi z}}{\Gamma(a-z) }
{\prod_{k=1}^{N+1}  \Gamma(\alpha_k - z) \prod_{j=1}^N\Gamma(z+\beta_j)}.
\end{align}
The factor $\prod_{k<j} (z_k-z_j) \Big(\tan z_j-\tan z_k\Big)$ is given by the product of two  Vandermonde determinants
\begin{align}
\label{Vandermonde2}
(-1)^{N(N-1)/2}\,\,
\det \begin{vmatrix}
1&\cdots&1\\
z_1  &\cdots&z_N
\\
\vdots & \ddots &\vdots
\\
z_1^{N-1}&\cdots&z_N^{N-1}
\end{vmatrix}
\times\det \begin{vmatrix}
1&\cdots&1\\
t_1  &\cdots&t_N
\\
\vdots & \ddots &\vdots
\\
t_1^{N-1}&\cdots&t_N^{N-1}
\end{vmatrix},
\end{align}
where $t_k\equiv \tan\pi z_k$. Taking into account the symmetry of the integrand under permutations of the arguments $z_1,\ldots, z_N$ one
can represent the integrals~\eqref{RR2} as follows
\begin{flalign}\label{ww}
\mathscr R_\pm(a,\alpha,\beta) & = (-\pi)^{\frac{N(N-1)}2} N!\int_{-i\infty}^{i\infty}\cdots\int_{-i\infty}^{i\infty}
\prod_{k=1}^N z_k^{k-1}\big(\cos\pi z_k\big)^{N-1} Q_\pm(z_k)
\det |\mathrm T_{N}|
\prod_{\ell=1}^N\frac{dz_\ell}{2\pi i},
&
\end{flalign}
where  the matrix $\mathrm T_{N}$  is the matrix of tangents in~\eqref{Vandermonde2}. Since the entries of the $k$-th column of the matrix
$\mathrm T_N$ depend only on one variable, $z_k$,  one obtains the following  (determinant) representation for the integrals in question
\begin{align}
\mathscr R_\pm(a,\alpha,\beta)= (-\pi)^{N(N-1)/2} N!\det| \mathcal Q_\pm|.
\end{align}
Here $\mathcal Q_\pm$ are  $N\times N$ matrices with the entries
\begin{align}\label{Qpm}
 (\mathcal Q_\pm)_{mk}=\frac1{2\pi i}\int_{-i\infty}^{i\infty} dz \, (\cos\pi z)^{N-1}\,
  (\tan\pi z)^{m-1} \,z^{k-1}\, Q_\pm(z).
\end{align}
 Let  $I^+_{mk}(z)$ and $I^-_{mk}(z)$ be the integrands in this expression. It is easy to see that
 $I^+_{mk}(iu)$ and $I^-_{mk}(-iu)$ decay exponentially $(\sim e^{-2\pi u})$ when
$u\to\infty$ and power-like when $u\to -\infty$. The integrands in the expression~\eqref{Qpm},
 $I^+_{mk}(iu)$ and $I^-_{mk}(-iu)$, decay exponentially $(\sim e^{-2\pi u})$ when
$u\to\infty$ and power-like when $u\to -\infty$
\begin{align}\label{IIpower}
I_{mk}^{\pm}(\mp iu)&\underset{u\to\infty}{=} 2\pi^N u^{\mathscr A+\mathscr B-a-N+k-1} \times (\mp i)^{\mathscr B-\mathscr A+a+m+k-2}
\left(1+O\left(1/u\right)\right).
\end{align}
It is easy to see from Eqs.~\eqref{Qpm} and \eqref{IIpower} that all matrix elements in the last row,  $ (\mathcal Q_\pm)_{mN} $, diverge
when $a\to \mathscr A+\mathscr B$. Namely,
\begin{align}
(\mathcal Q_\pm)_{mN}\underset{a\mapsto \mathscr A+\mathscr B}{=}
    \pi^{N-1} \frac{e^{\mp i\pi \mathscr B}(\mp i)^{m+N-2}}{a-\mathscr A-\mathscr B}+O(1),
\end{align}
while the elements $(\mathcal Q_\pm)_{m,k}$ with $k<N$ are finite.
 Thus  the singular part of the integrals $\mathscr R_\pm$ in the limit $a\to \mathscr A+\mathscr B$ can be represented in the following form
\begin{flalign}\label{rrr}
\mathscr R_\pm(a,\alpha,\beta) &\underset{a\mapsto \mathscr A+\mathscr B}{=}
(-\pi)^{\frac{N(N-1)}2} N!
\frac{e^{\mp i\pi \mathscr B}(\mp i\pi)^{N-1}}{a-\mathscr A-\mathscr B}\times
\notag\\
&\qquad\int_{-i\infty}^{i\infty}\cdots\int_{-i\infty}^{i\infty}
\left(\prod_{k=1}^{N-1} z_k^{k-1}\big(\cos\pi z_k\big)^{N-1} Q_\pm(z_k)\right)
\det |\widehat {\mathrm T}^\pm_{N}|
\prod_{\ell=1}^{N-1}\frac{dz_\ell}{2\pi i} + O(1),
&
\end{flalign}
where the matrices $\widehat{\mathrm T}^\pm_{N}$ read
\begin{align}
\widehat{\mathrm T}^\pm_{N} &=
 \begin{pmatrix}
1&1&\cdots& 1\\
t_1 & t_2 &\cdots&(\mp i)
\\
\vdots & \ddots &\vdots
\\
t_1^{N-1}& t_2^{N-1}&\cdots&(\mp i)^{N-1}
\end{pmatrix}.
\end{align}
The determinants of these matrices   can be written as follows
\begin{align}\label{III}
\det\widehat{\mathrm T}^{\pm}_N & =  (-1)^{N-1} \prod_{k=1}^{N-1}(t_k\pm i) \times\det \mathrm T_{N-1},
\end{align}
where $ \mathrm T_{N-1}$ is the $(N-1)\times(N-1)$ Vandermonde marix
\begin{align}
 \mathrm T_{N-1} &= \begin{pmatrix}
1&1&\cdots& 1\\
t_1 & t_2 &\cdots& t_{N-1}
\\
\vdots & \ddots &\vdots&\vdots
\\
t_1^{N-2}& t_2^{N-2}&\cdots& t_{N-1}^{N-2}
\end{pmatrix}\,.
\end{align}
Since $$t_k\pm i=\pm i \frac{e^{\mp i \pi z_k}}{\cos\pi z_k}$$  one can rewrite Eq.~\eqref{rrr} as follows
\begin{flalign}\label{RR3}
\mathscr R_\pm(a,\alpha,\beta) &\underset{a\mapsto \mathscr A+\mathscr B}{=}
\frac{e^{\mp i\pi \mathscr B}}{a-\mathscr A-\mathscr B}{(-\pi)^{(N-1)(N-2)/2} N!}\times
\notag\\
&\qquad \int_{-i\infty}^{i\infty}\cdots\int_{-i\infty}^{i\infty}
\left(\prod_{k=1}^{N-1} z_k^{k-1}\big(\cos\pi z_k\big)^{N-2} \widetilde Q_\pm(z_k)\right)
\det |\mathrm T_{N-1}| \prod_{\ell=1}^{N-1} \frac{dz_\ell}{2\pi i}  + O(1), &
\end{flalign}
where the functions $\widetilde Q_\pm$ are given by the product of gamma functions
\begin{align}
\widetilde Q_\pm(z)&=e^{\mp i \pi z}Q_\pm(z)\Big|_{a=\mathscr A+\mathscr B}
=
\frac{1}{\Gamma(\mathscr A+\mathscr B-z) }
{\prod_{k=1}^{N+1}  \Gamma(\alpha_k - z) \prod_{j=1}^N\Gamma(z+\beta_j)}.
\end{align}
Since the integral in \eqref{RR3} coincides, up to  changes $N\to N-1$ and $Q\to\widetilde Q$,   with the integral~\eqref{ww} it can be
written in the form of the  MB integral
\begin{align}\label{residue2}
\mathscr R_\pm &\underset{a\mapsto \mathscr A+\mathscr B}{=}{N}\frac{e^{\mp i\pi \mathscr B}}{a-\mathscr A-\mathscr B} \times \mathscr T_{N-1}
+\text{regular terms},
\end{align}
where
\begin{align}
\mathscr T_{N-1} &=
\int_{-i\infty}^{i\infty}\!\!\!\cdots \int_{-i\infty}^{i\infty} \!
\frac{\prod_{j=1}^{N-1}\prod_{k=1}^{N+1}\Gamma(\alpha_k-z_j)
\prod_{m=1}^N\Gamma(z_j+\beta_k)}{\prod_{k=1}^{N-1} \Gamma({a}-z_k)
\prod_{k < j} \Gamma(z_k-z_j) \Gamma(z_j-z_k)}\prod_{k=1}^{N-1} \frac{dz_k}{2\pi i} %
\end{align}
and $a=\mathscr A+\mathscr B$.

Comparing residues at $ a=\mathscr A+\mathscr B$ on the both sides of \eqref{Rpm1} one gets the following equation
\begin{align}
\mathscr T_{N-1}=(N-1)!\frac{\prod_{k=1}^{N}\prod_{j=1}^{N+1}\Gamma(\alpha_j+\beta_k)}{\prod_{j=1}^{N+1}\Gamma\left( \mathscr A+\mathscr
B-\alpha_j\right)},
\end{align}
which is, up to renumeration $N\to N-1$, nothing but the second Gustafson integral~\eqref{Gustafson-IA}.

\section{Summary}

The Mellin - Barnes integrals and their $q$- and elliptic generalizations play an important role in many topics in physics and mathematics,
see e.g. refs.~\cite{SpiridonovEssays,MR1694379,MR2434345}. Many important results generalizing first and second Barnes' lemmas to
certain Lie algebras were obtained by R.A.~Gustafson~\cite{Gustafson92,Gustafson94}. In this article, we have presented an alternative
proof of the integral~\eqref {Gustafson-IA}, which is based on the study of the analytic properties of some auxiliary integrals as
functions of external parameters. This technique is not restricted to the case under consideration. It has  already been applied to the
complex $\mathrm{SL}(2,\mathbb C)$ MB integrals, see ref.~\cite{DM2020}. Finally we note that the same analysis, applied to the MB integral
\begin{flalign*}\label{three-stars}
\int_{-i\infty}^{i\infty}\!\!\!\cdots\int_{-i\infty}^{i\infty}&
\frac{\prod_{k=1}^{2N+1} \prod_{j=1}^{N} \Gamma(\alpha_k \pm z_j)}{\prod_{k=1}^N\Gamma(\beta\pm z_k)\Gamma(\pm 2z_k)
\prod_{k< j}\Gamma(\pm z_k \pm z_j)}
\prod_{n=1}^{N}  \frac{dz_n}{4\pi i}
=
 N!\Gamma(\beta-A)\frac{\prod_{k< j}\Gamma(\alpha_k+\alpha_j)}{\prod_{k=1}^{2N+1} \Gamma(\beta-\alpha_k)}, &
\end{flalign*}
where $\Gamma(\pm a)=\Gamma(a)\Gamma(-a)$, $\Gamma(a\pm b)=\Gamma(a+b)\Gamma(a-b)$, etc., allows one to obtain the MB integral
~\cite[Theorem 5.3]{Gustafson92}.

\subsection*{Acknowledgements}
The author is grateful to Sergey Derkachov for useful discussions. This work was supported by the Russian Science Foundation project No
19-11-00131 and by the DFG grants MO 1801/4-1, KN 365/13-1.

\end{document}